\documentstyle[11pt,newpasp,twoside]{article}
\input epsf

\begin{document}

\title{Measurements of Tidal Tail Populations Designed to Optimize
Their Use as Potential Probes}

\author{Kathryn V. Johnston}
\affil{Van Vleck Observatory, Wesleyan University, Middletown,
CT 06457}

\begin{abstract}
The launching of up to four astrometric missions in the
next decade will enable us to make measurements of stars
in tidal streamers from Galactic satellites of sufficient accuracy
to place strong constraints on the mass distribution in the Milky Way.
In this paper we simulate observations of debris populations in order
to assess the required properties
of any data set chosen to implement this experiment.
We apply our results to find the desired target properties of
stars (e.g., accuracy of velocity and proper motion measurements)
associated with the dwarf spheroidal 
satellites of the Milky Way.
\end{abstract}

\section{Introduction}

Stars lost from a satellite galaxy
in orbit around the Milky Way will gradually drift
ahead or behind their parent system to form streams of debris along
its orbit (Tremaine 1993; Johnston 1998).
Assuming we know the phase-space coordinates of the parent satellite,
the unique properties of tidal debris streams can be exploited to
measure the potential of the Milky Way in two ways: \\
{\bf MD Algorithm:} 
Integrate the satellite's orbit backwards and forwards
and try to fit any available observations 
to the orbit, with
the ``best'' potential corresponding to the closest fit
(e.g., Murali \& Dubinski 1999). \\
{\bf JZSH Algorithm:} Integrate the stars and satellite backwards 
and choose the ``best'' potential to be the one in which most
stars recombine with the satellite
(e.g., Johnston et al. 1999, hereafter JZSH99). \\
The MD algorithm can be applied with limited observations 
(e.g., just angular positions
and line-of-sight velocities) of any accuracy.
However, in reality debris in trailing/leading
streamers are {\it not} on exactly the same orbit
as their parent satellite, but rather in orbits
offset in energy from the satellite's orbit
by $\sim \pm \epsilon$ where
\begin{equation}
	\epsilon=r_{\rm tide} {\partial\Phi \over \partial r}\approx s v_{\rm circ}^2.
\label{eps}
\end{equation}
Here, $r_{\rm tide}=rs$ is the tidal radius of the satellite,
$\Phi({\bf r})$ is the Galactic potential,
$v_{\rm circ}$ is the circular velocity of the Milky Way
and $s$ is the {\it tidal scale}
\begin{equation}
	s=\left({m_{\rm sat} \over M_{\rm Gal}}\right)^{1/3}, 
\label{tfac}
\end{equation}
where $m_{\rm sat}$ is the satellite's mass and $M_{\rm Gal}$ is the mass of the Galaxy 
enclosed within the satellite's orbit.
Hence, once the observational errors are small enough to detect the
offset of the particles from the satellite's orbit
the MD algorithm will be biased by this systematic offset, rather
than limited by the errors themselves.
In contrast, the JZSH algorithm exploits the fact that debris stars
are on orbits with different time periods than the satellite's
own and hence will {\it only}
work with observations of five of the phase-space coordinates of sufficient
accuracy that the offset from the orbit can be detected
(the sixth coordinate, if poorly known,
can always be found by applying the principle
of energy conservation).
The dividing line between useful application of
these two approaches will therefore occur where the velocity measurement errors
$\Delta v$ satisfy
\begin{equation}
	v_{\rm circ} \Delta v \approx \epsilon 
	\Rightarrow {\Delta v \over v_{\rm circ}} \approx s.
\label{verror}
\end{equation}

Once measurements of proper motions and velocities of
debris can be made with accuracies 
of order 3-30 km/s the second approach will be more accurate 
for all Milky Way satellites (depending on the mass and orbit of
the satellite --- see Table 1)
Such accuracies will become possible with the next generation astrometric
missions: NASA's 
Space Interferometry Mission (SIM) 
and 
ESA's Global Astrometric Imager for Astrophysics (GAIA).

JZSH99 have already demonstrated that the JZSH algorithm can in principle recover
both the mass and geometry of the Milky Way with few percent
accuracies using just 100 stars.
In this paper, we examine how these uncertainties
grow with the limitations likely to be present in a real data set.

\section{Methods}

We test how well we can recover the Milky Way
potential by ``observing''
debris populations at the end of simulations of satellite destruction and
then applying the JZSH algorithm to our data sets.
We illustrate our results in this paper with tests on
simulations referred to as Models 1/2/3/4,
which had satellite masses 
5.1e6/2.9e7/6.9e7/1.0e8 $M_\odot$,
on orbits with pericenters at 43/43/26/30 kpc and radial time periods of
2.5/2.5/2.0/1.0 Gyears.
These parameters corresponded to tidal scales
$s=0.026/0.046/0.075/0.081$ (see equation [3]). 
The simulation technique and Galactic  model are
described in Johnston, Spergel \& Hernquist (1995).

We restrict ourselves to asking how well we can recover the
circular velocity of the halo component that was used in the
simulations, assuming we know the other parameters in the
potential.
As an an indication of the error, we attempt to recover $v_{\rm circ}$
from ten data sets, each containing
a different group of particles or a different error realization.
We take the ``error'' to be the
dispersion in the recovered $v_{\rm circ}$
in units of the true circular velocity ($=\sigma_{v_{\rm circ}}/v_{\rm circ}$). 

\begin{figure}[t]
\plotone{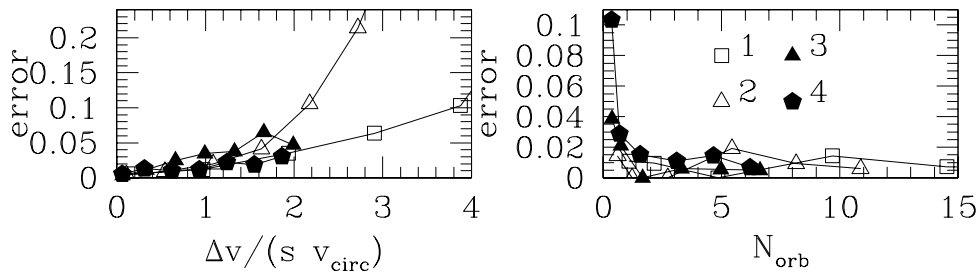}
\caption{Fractional error in $v_{\rm circ}$ recovered as a function
of velocity measurement errors (left hand panel) and number
of orbits experienced by debris star (right hand panels)
\label{fig1}}
\end{figure}

\section{Results}

{\bf Measurement Errors of Debris Stars}
As noted in \S 1, we expect the JZSH algorithm 
to require knowledge of velocities
such that $\Delta v/v_{\rm circ} < s$.
To test this, we observed 128 stars in Models 1-4. We then added 
errors chosen from gaussians of mean zero and dispersions 
$\Delta v$ and $\Delta v/d$  
to each line of sight velocity and
each component of proper motion respectively, and ran the JZSH algorithm
to recover $v_{\rm circ}$. 
This was repeated 10 times with the same sample of stars (but different
error realizations) for 
$\Delta v=1,5,10,15,20,25,30$km/s and the dispersion among
the results from the 10 realizations calculated. The errors
found, shown in the left hand panel of
Figure 1, confirm the intuition outlined in \S 1.

{\bf Length of Tidal Streamers}
A typical debris star will drift
by \\
$\Delta \Psi=\epsilon(d\Omega/d\epsilon)(2\pi/\Omega)\sim
r_{\rm tide}(d\Omega/dr)(2\pi/\Omega)$ in angle away
from the satellite in one orbit (where $\Omega$ is the angular velocity
of the satellite), so a star
with separation $\Delta \Psi$
from the satellite, was most likely lost $N_{\rm orb}$ orbits ago, where
\begin{equation}
	N_{\rm orb} 
	\sim {\Delta \Psi \over 2\pi}
{\Omega \over r_{\rm tide}} {dR \over d\Omega}
	\sim {\Delta \Psi \over 2\pi} {1 \over s}.
\end{equation} 
To test how the length of the tidal streamer sample might
affect the accuracy of the recovered potential
we observed 10 random samples of
128 particles in each simulation, with the samples chosen to be within
$\Delta \Psi=10, 20, 45, 90, 135$ and 180 degrees
of the satellite.
The right hand panel of Figure 1 plots the results for each simulation
as a function of $N_{\rm orb}$ defined from $\Delta \Psi$ via 
equation (4).
The plot suggests that if we find $\sim 100$ stars in a streamer
we would like to observe them
at sufficiently large angle from the satellite
to be sensitive to 1-2 orbits.

{\bf Knowledge of Satellite Distance}
We anticipate knowing the line-of-sight velocity and proper motion of
the satellite with a precision at least comparable to 
those of the debris stars. However, the satellite's distance $r_{\rm sat}$
may be poorly known (or unknown). This means that though
we can expect to place strong limits on the mass distribution in
the Galaxy, $M_{\rm Gal}({\bf \rm r}/r_{\rm sat})$,
the physical scale of
that distribution will be dependent on our measurement of
$r_{\rm sat}$ and uncertain at the same level.

{\bf Dynamical Friction}
Binney \& Tremaine (1987) 
show that the frictional force on a satellite on a circular orbit
in an isothermal sphere is given by (see their equation [7.23])
$F=-0.428 ln \Lambda {G m_{\rm sat}^2/ R^2},$
where $ln \Lambda$ is the Coulomb Logarithm.
The change in orbital energy per orbit will be
$\Delta E_{\rm orb}^{\rm df}=
F (2\pi R)/m_{\rm sat}$, so
\begin{equation}
	{\Delta E_{\rm orb}^{\rm df} \over \epsilon} 
		=-0.428 ln \Lambda {G m_{\rm sat} \over R^2} 2\pi R 
			{1 \over v_{\rm circ}^2 s} 
		\approx-2.69 ln \Lambda s^2
\label{df}
\end{equation}

{\bf Evolution of the Galactic Potential}
The properties of debris trails will also be affected by
the time-dependence of the Milky Way's potential (e.g.,
Zhao et al. 1999).
However, the existence of the $> 7$Gyear old thin disk suggests
that potential evolution has not been significant within the
the last $N_{\rm orb}=1$ orbit of any of the Galactic satellites.

\begin{table}[t]
\begin{center}
\begin{tabular}{lrcccrrrc}
Name 	& $R$	& dist & $m_{\rm sat}$ & $s$ & $\Delta v$ & $\Delta \mu$& $\Delta \Psi$ 
& ${\Delta E_{\rm orb}^{\rm df} \over \epsilon}$ \\
	& (kpc)	& mod  & $M_{\odot}$	&& (km/s)& $\mu$as/yr & (degrees)  \\
\hline
\hline
Sgr   &   16.0 & 16.9 & 5.0E+08 & 0.151 &  30.3 & 252 & 54 &  0.393 \cr
Sex   &   86.0 & 19.7 & 2.6E+07 & 0.032 &   6.5 &  15 & 11  &  0.026 \cr
Car   &   86.6 & 19.7 & 1.1E+07 & 0.024 &   4.8 &  11 & 9  &  0.016 \cr
LeoII &  210.0 & 21.6 & 1.4E+07 & 0.020 &   3.9 &   4 & 8  &  0.010 \cr
\end{tabular}
\end{center}
\caption{Properties of tidal tail data sets
associated with each of Milky Way's satellites needed to
apply the JZSH algorithm.
\label{satstab}
}
\end{table}

\section{Conclusion}

Table 1 lists the required accuracy of
line-of-sight velocity and proper motion measurements
of tidal stream stars
(columns 6 and 7), length of tidal stream population (column 8) and
effect of dynamical friction (column 9) for a sample of Galactic satellites.
The numbers demonstrate that the necessary
measurement accuracies are within
the capabilities of SIM and GAIA and that samples extending tens
of degrees along a streamer to select targets will be sufficient.
Dynamical friction needs only to be modeled carefully for the
innermost satellites.
We conclude that the JZSH algorithm, combined with data from 
SIM or GAIA will
significantly improve our knowledge of the Galactic potential.

\acknowledgements
I thank David Spergel for helpful comments on this manuscript.
This work was supported in part by NASA grant NAG5-9064.


\begin{references}
\reference
Binney, J. \& Tremaine, S. 1987, {\it Galactic Dynamics} (Princeton 
Uni. Press)
\reference
Johnston, K.V. 1998, \apj, 495, 297
\reference
Johnston, K. V., Spergel, D. N. \&
Hernquist, L. 1995, \apj, 451, 598
\reference
Johnston, K. V., Zhao, H.S., Spergel, D. N.  \& Hernquist, L. 1999 (JZSH)
\reference
Murali, C. \& Dubinski, J. 1999, \aj, 118, 921
\reference
Tremaine, S. 1993 in {\it Back to the Galaxy} eds. S. S. Holt \& F. Verter,
(AIP Conf. Proc. : New York), p. 599
\reference
Zhao, H.S., Johnston, K.V., Spergel, D.N. \& Hernquist, L., 1999, 
\end{references}
\end{document}